\documentclass{ws-p8-50x6-00}

\newcommand{\vc}[1]{{\vec {#1}\,}}

\begin{document}

\title{Transverse Energy Production at RHIC}

\author{Qun Li$^1$, Yang Pang$^{1,2}$, Nu Xu$^1$}

\address{$^1$70-319, Lawrence Berkeley National Laboratory\\
1 Cyclotron Road,Berkeley, CA 94720, USA\\
$^2$Department of Physics, Columbia University, New York, NY 10027, USA\\
E-mails: qli@lbl.gov, nxu@lbl.gov, ypang@lbl.gov}


\maketitle

\abstracts{We study the mechanism of transverse energy ($E_T$) production in
Au+Au collisions at RHIC.  The time evolution starting from the initial energy loss to the 
final $E_T$ production is closely examined in transport models. The relationship 
between the experimentally measured $E_T$ distribution and the maximum energy density
achieved is discussed.}

\section{Introduction}
The Relativistic Heavy Ion Collider (RHIC) 
will provide us with a unique opportunity to study matter under extreme
conditions. The transition
from hadronic degrees of freedom to quark and gluon degrees of freedom is
expected from collisions at these energies.
The transverse energy spectrum will be among the very first results from RHIC.
One important question we would like to answer from the day-one physics at RHIC is the
maximum energy density reached in central Au+Au collisions. 
Lattice QCD~\cite{lattice} calculations give us the critical energy density for the
transition to quark-gluon plasma. 
In this talk, we explore the relationship between the transverse energy spectrum
and the energy density, and discuss the possibility of experimentally
determine the maximum energy density.

A general framework for computing and analysing energy-momentum tensor is introduced. 
As the first part of this study, we construct the energy-momentum tensor
for simulated events from transport model RQMD~\cite{rqmd}. We also find the 
transverse energy, the equation of state, and the energy density for these
events. Currently we are using our techniques to study the parton cascade model
VNI~\cite{vni} as well as other event generators. 

\section{Energy-Momentum Tensor in Transport Models}
In transport model, the energy-momentum tensor
\begin{equation}
 T^{\mu\nu}(\vec x\,,t) \equiv \sum_i\int
{d^3p}\frac{p^{\mu}p^{\nu}}{p^0}f_i(\vec x\,,\vec p\,,t) \ ,
\end{equation}
and the particle number current
\begin{equation}
j^\mu(\vc x, t) \equiv \sum_i\int {d^3p\over p^0} p^\mu f_i(\vc x,\vc p, t) \ .
\end{equation}
where $f_i(\vec x,\vec p,t)$ is the distribution functions for particle type
$i$. 

In a relativistic cascade, each particle is represented by 
a point in both position and momentum space, and
the distribution function is an ensemble average of the
$\delta$-functions,
\begin{equation}
f_i(\vec x\,,\vec p\,,t)= \left\langle \sum_k\delta^3(\vec x-\vec r_k(t)) 
\delta^3(\vec p-\vec p_k(t)) \right\rangle \ .
\end{equation}
The energy-momentum tensor and the particle current become
\begin{equation}
T^{\mu\nu}(\vc x, t) 
= \lim\limits_{V\to 0}\frac{1}{V}\left<\sum\limits_k^{\vc x_k(t)\in V}
{p^\mu_k p^\nu_k \over p^0_k}\right> \ ,
\end{equation}
and
\begin{equation}
j^\mu(\vc x, t) = \lim\limits_{V\to 0}
  \frac{1}{V} \left<\sum\limits_k^{\vc x_k(t)\in V} {p^\mu_k\over p^0_k}\right> \ .
\end{equation}

At any given space-time location, $T^{\mu\nu}$, being a symmetric tensor,
has ten independent elements from which we can obtain ten physical
quantities: local energy density $\epsilon$, local pressures $\mathcal{P}_1,
\mathcal{P}_2, \mathcal{P}_3$, the flow velocity $\vec v_f$, and the
orientation of 
the principal momentum axises. Following the convention of 
Landau and Lifshitz~\cite{landau}, the local rest frame is defined as the
frame in which 
\begin{equation}
T^{\mu\nu} = \left(
\begin{array}{cc}
\epsilon & 0\\
0        & T^{ij}\\
\end{array}
\right)\ .
\end{equation}
The Lorentz boost of $T^{\mu\nu}$ to the above form gives us the flow
velocity. The momentum tensor $T^{ij}$ can be diagonalized by performing a
rotation. After the boost and the rotation~\cite{koch}, 
\begin{equation}
T^{\mu\nu} = \left(
\begin{array}{cccc}
\epsilon & 0&0&0\\
0        & \mathcal{P}_1 &0 &0\\
0        &0  & \mathcal{P}_2&0\\
0        &0 &0& \mathcal{P}_3 \\
\end{array}
\right) \ .
\end{equation}

The local particle density
\begin{equation}
\rho \equiv j^\mu u_\mu \ ,
\end{equation}
where $u^\mu = (\gamma, \gamma\vc v)$ is the velocity four vector. We further
define 
\begin{equation}
{{\mathcal{P}}} \equiv (\mathcal{P}_1 + \mathcal{P}_2 + \mathcal{P}_3)/3
\ \ \ \ {\rm and} \ \ \ \ T \equiv \mathcal{P}/\rho \ .
\end{equation}
If there is local thermal equilibrium, then $\mathcal{P}$ would be the pressure
and $T$ would correspond to the temperature.

\section{Event Generator Studies}

\begin{figure}[tbp]
\begin{center}
\hbox{
\hspace{-0.25in}
\psfig{figure=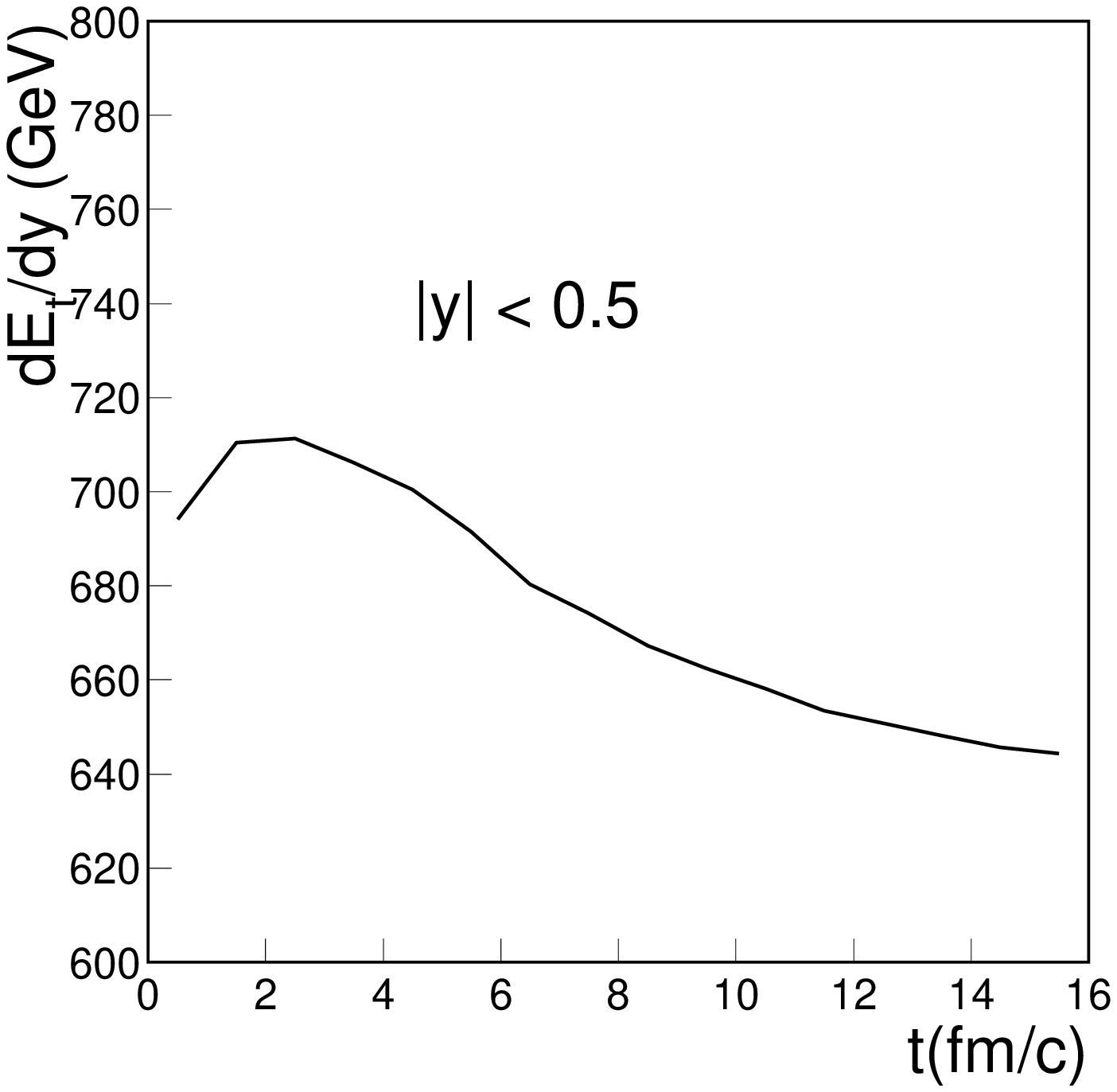,width=2.5in}
\hspace{0.05in}
\psfig{figure=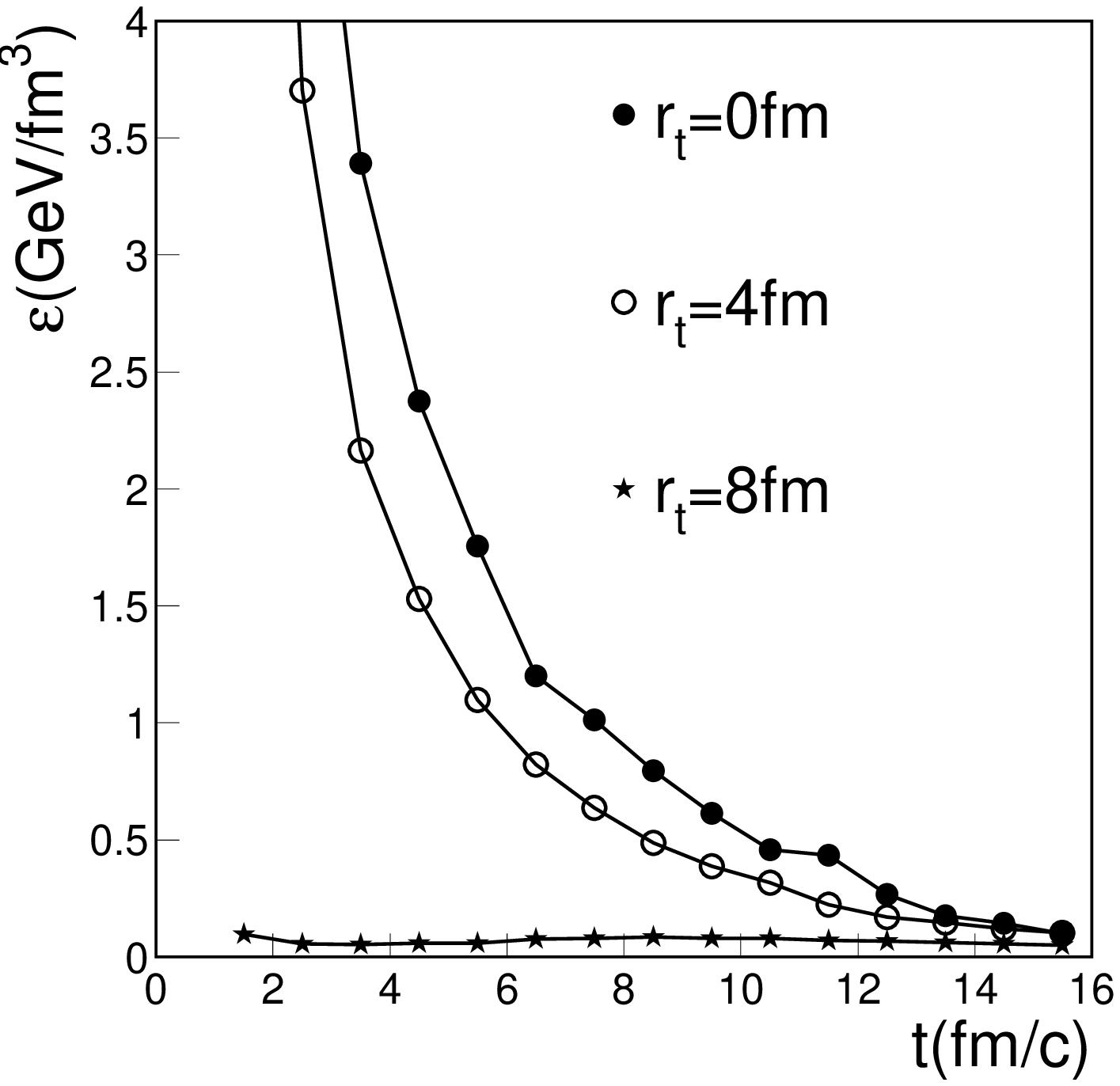,width=2.5in}
\hspace{-0.25in}
}
\hbox to \textwidth{\hspace{0.8in}(a) \hss (b)\hspace{0.8in}}
\end{center}
\caption{(a) Transverse energy at mid-rapidity as a function of time for central ($b=0$) Au+Au 
collisions at RHIC from RQMD v2.4; (b) Local energy density, defined by Eq.(6), as a function of
time at various positions ($z=0$ and $r_t=0,4,8$fm) for the same set of collisions.}
\label{fig1}
\end{figure}

\begin{figure}[tbp]
\begin{center}
\hbox{
\hspace{-0.25in}
\psfig{figure=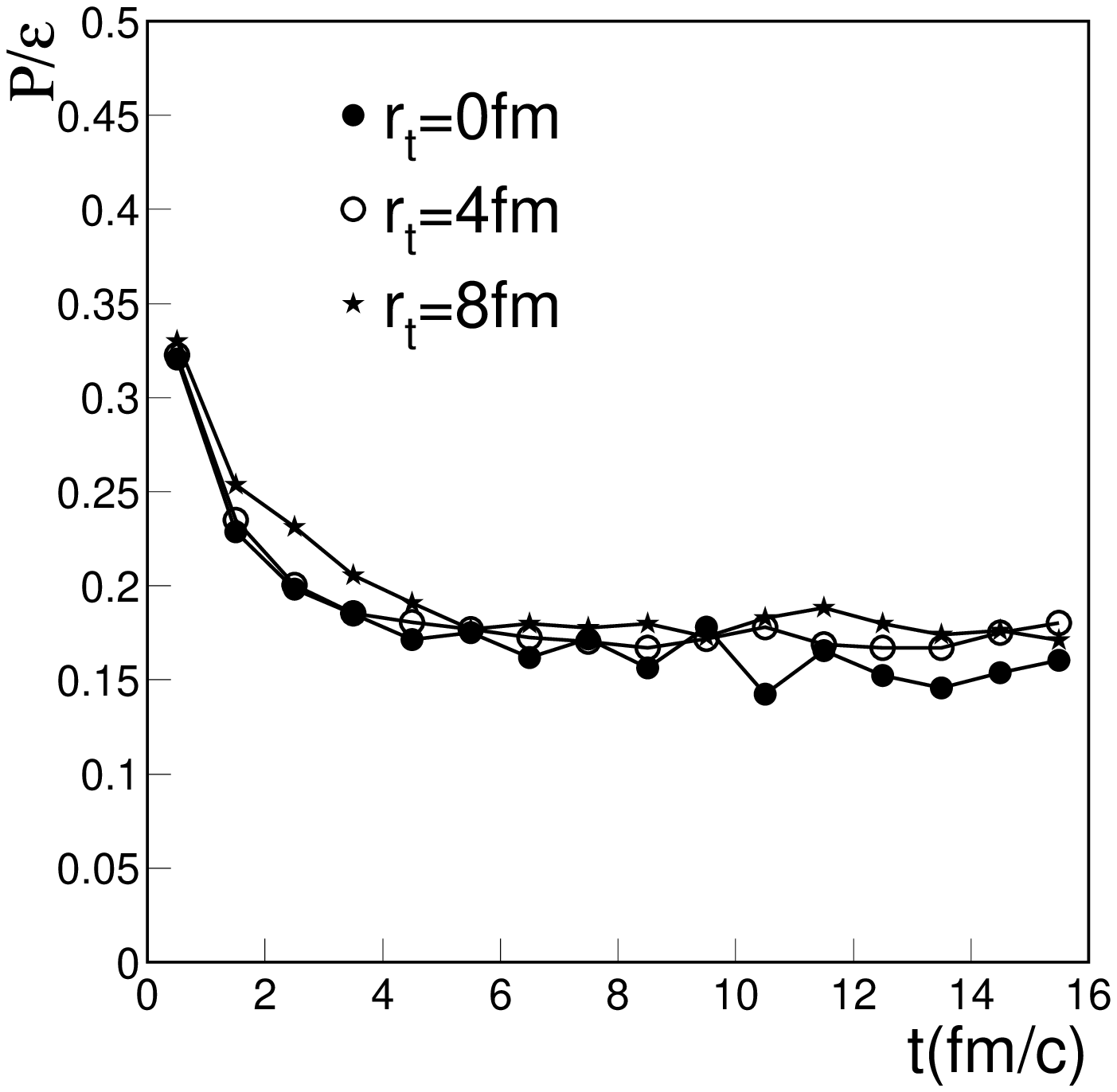,width=2.5in}
\hspace{0.05in}
\psfig{figure=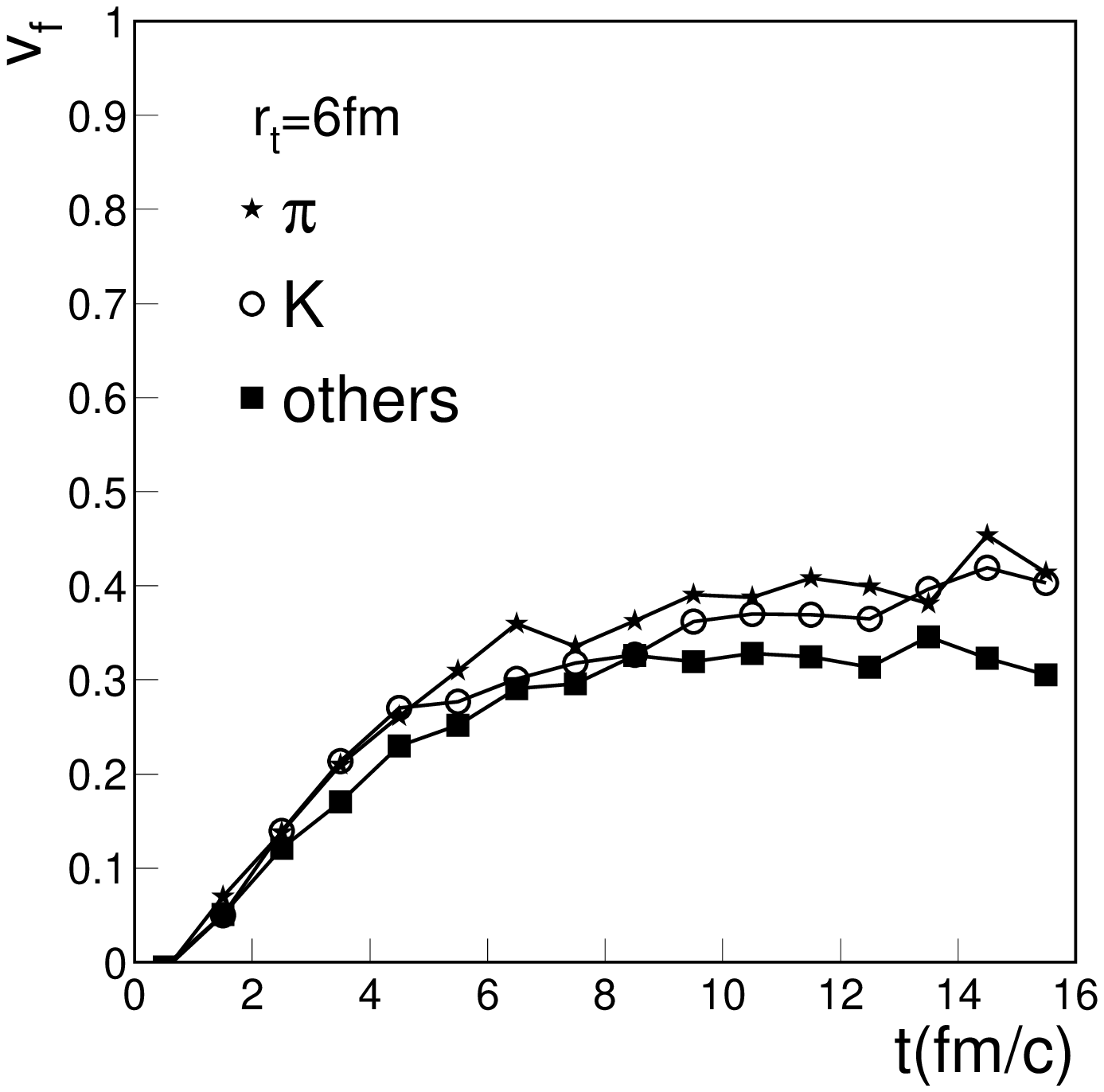,width=2.5in}
\hspace{-0.25in}
}
\hbox to \textwidth{\hspace{0.8in}(a) \hss (b)\hspace{0.8in}}
\end{center}
\caption{For the same set of collisions from RQMD v2.4 as in Figure 1 :(a) The ratio of 
local pressure,defined by Eq.(9),to local energy density as a function of time at 
various positions ($z=0$ and $r_t=0,4,8$fm); 
(b) Flow velocities for $\pi, K$, and for other particles as a function of time at 
$z=0$ and $r_t=6$fm.}
\label{fig2}
\end{figure}
There are, in general, two stages in the evolution of $E_t$. Initially, $E_t$ 
increases due to the transverse excitations from interactions. Afterwards, the longitudinal 
expansion results in a decreasing $E_t$. What we measure in the experiment is the final
$E_t$ at the end of this expansion. The details of the $E_t$ evolution, like all other 
observables in the relativistic heavy ion collisions, is model dependent. In order to get a 
reasonable estimate of the energy density through measurement, we need a systematic 
study  of all plausible models.  The transverse energy $E_t$ and the energy density $\epsilon$
 from the transport model RQMD is shown in Figure 1. 

A total of 144 central ($b=0$) Au+Au events, with full event histories, at RHIC 
energy (100~GeV/A+100~GeV/A) was generated using RQMD version $2.4$~\cite{rqmd}. 
The energy-momentum tensor and the particle current is computed according to 
Eqs. (4) and (5), where the 
small volume V is chosen to be a sphere of an 1~fm radius. The local energy density 
$\epsilon$, pressure $\mathcal{P}$, and the flow velocity $\vec v_f$ are obtained 
from Eqs. (6)-(9). Figure 2 shows the variation of ${\mathcal{P}}/{\epsilon}$ and the 
development of transverse flow velocity.

In the RQMD events, during the early stage, the longitudinal pressure is higher than the 
transverse pressure. After about 7~fm/c, the system is approximately isotropic, but it is 
still not in a local thermal equilibrium. Figure 2(b) displays one such nonthermal 
behaviour in the flow velocity distribution. Particles with lighter masses, e.g. pions
and Kaons, have greater collective velocities than heavier particles, i.e. there is no common 
flow velocity.

Clearly, the system in Figures 1 and 2, can not be described by a simple Bjorken expansion picture. 
There are expansions along both the longitudinal and the transverse directions, and the system 
is not in a local thermal equilibrium. 
The time evolution of $\epsilon$ and $E_t$, in Figure 1, are not exactly correlated. So the 
measurement of $E_t$ by itself is not sufficient to determine $\epsilon$.

The initial state in RQMD is not partonic, so it may not be applicable for RHIC. But the model may 
still have a reasonable parametrisation for stopping and for initial transverse energy production, 
and it has a well tested hadronic final state after-burner. Models like VNI, HIJING, VENUS, 
and UrQMD have quite different descriptions of the initial stages at RHIC. By coupling 
these initial conditions with a common hadronic after-burner, we will have a better understanding 
of the relationship between the observables and the maximum energy density.

\section{Remarks}
An estimate of the maximum energy density might be possible if,in addition to transverse energy,
we also have a good measurement of several collective observables such as the radial and 
elliptic flows, the system size from HBT, 
and other correlations . This kind of estimate would still be model 
dependent, so we need to make a survey of all models.

\section*{Acknowledgements}
We wish to thank H.~Sorge for helping us with the intermediate
states of RQMD events. We also wish to thank Dr. M.~Gyulassy, Dr. S.~Jeon, 
and Dr. S.~Voloshin for many fruitful discussions, and thank Dr. G.~Baym for
suggesting this study. The work is supported in part by U.S. Department of Energy
under contract No. DE-AC03-76SF00098, No. FG02-92ER40699 and the Alred P. Sloan 
Foundation (Y.P).


\end{document}